\documentclass{IEEEcsmag}

\usepackage[colorlinks,urlcolor=blue,linkcolor=blue,citecolor=blue]{hyperref}

\usepackage{upmath}
\usepackage{amsmath}
\usepackage{amssymb}
\usepackage{pifont}
\usepackage{multirow}
\usepackage{balance}
\usepackage[caption=false]{subfig}
\jvol{XX}
\jnum{XX}
\paper{8}
\jmonth{March}
\jname{IEEE Multimedia}
\pubyear{2020}

\newcolumntype{P}[1]{>{\centering\arraybackslash}p{#1}}
\newcolumntype{M}[1]{>{\centering\arraybackslash}m{#1}}
\newcommand{\hxy}[1]{\textcolor{black}{#1}}
\begin{document}

\sptitle{Feature Article}

\title{Key-point Sequence Lossless Compression for Intelligent Video Analysis}

\author{Weiyao Lin \hspace{5cm} Xiaoyi He}
\affil{Shanghai Jiao Tong University \hspace{2.34cm} Shanghai Jiao Tong University}

\author{Wenrui Dai \hspace{5cm} John See }
\affil{Shanghai Jiao Tong University \hspace{2.35cm}  Multimedia University}

\author{Tushar Shinde \hspace{4.54cm} Hongkai Xiong }
\affil{Indian Institute of Technology Jodhpur \hspace{1.21cm} Shanghai Jiao Tong University}

\author{\hspace{0.1pt}Lingyu Duan}
\affil{Peking University}


\begin{abstract}
Feature coding has been recently considered to facilitate intelligent video analysis for urban computing. Instead of raw videos, extracted features in the front-end are encoded and transmitted to the back-end for further processing. In this article, we present a lossless key-point sequence compression approach for efficient feature coding. The essence of this predict-and-encode strategy is to eliminate the spatial and temporal redundancies of key points in videos. Multiple prediction modes with an adaptive mode selection method are proposed to handle key-point sequences with various structures and motion. Experimental results validate the effectiveness of the proposed scheme on four types of widely-used key-point sequences in video analysis.
\end{abstract}

\maketitle

\begin{figure*}[!t]
\centerline{\includegraphics[width=\textwidth]{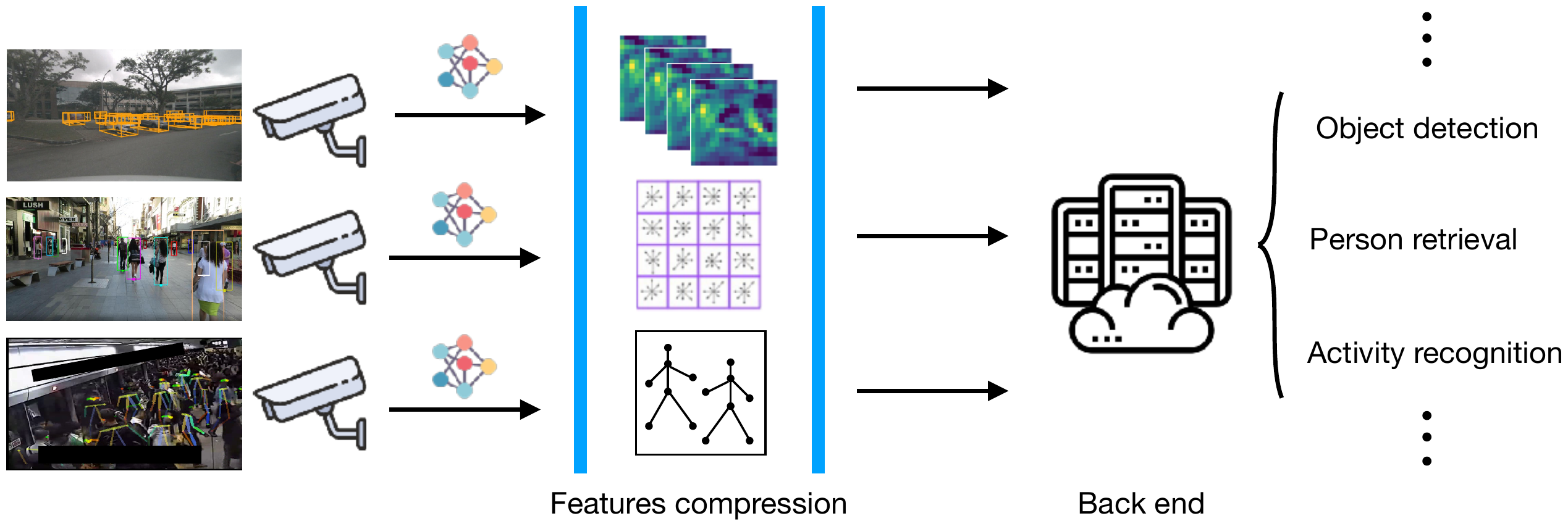}}
\caption{Illustration of the feature compression and transmission framework. \hxy{Best viewed in color.}}
\label{fig:frontback}
\end{figure*}

\chapterinitial{Intelligent video analysis,} involving applications such as activity recognition, face recognition and vehicle re-identification, has become part and parcel of smart cities and urban computing.
Recently, deep learning techniques have been adopted to improve the capabilities of urban video analysis and understanding by leveraging on large amounts of video data. With widespread deployment of surveillance systems in urban areas, massive amounts of video data are captured daily from front-end cameras. However, it remains a challenging task to transmit the large-scale data to the back-end server for analysis, although the state-of-the-art High Efficiency Video Coding (HEVC)~\cite{HEVC} and on-going Versatile Video Coding (VVC) standards present reasonably efficient solutions.
An alternative strategy that transmits the extracted and compressed compact features, rather than entire video streams, from the front-end to the back-end, is illustrated in Figure~\ref{fig:frontback}. These \emph{feature streams}, when passed to the back-end, enable various video analysis tasks to be achieved efficiently.
Here, we summarize the advantages of transmitting information via feature coding in a lossless fashion: 
(1) Lossy video coding would affect the fidelity of reconstructed videos and subsequent feature extraction at the back-end, which leads to degraded accuracy in video analysis tasks;
(2) Transmitting features rather than videos can mitigate privacy concerns to sensitive scenes such as in hospitals and prisons;
(3) Computational balance can be struck between the front-end and back-end processing, as decoded features are directly utilized for analysis in the back-end.

In video analysis, common features include hand-crafted features (e.g., LoG and SIFT descriptors), deep features and other contextual information (e.g., segmentation information, human and vehicle bounding boxes, facial and body pose landmarks). Among these features, key-point sequence is one of the most widely used type of feature. Key-point information like facial landmarks, human body key-points, bounding boxes of objects and region-of-interests (ROIs) for videos are essential for many applications, e.g., face recognition, activity recognition, abnormal event detection, and ROI-based video transcoding.

\begin{figure}[!t]
\centerline{\includegraphics[width=\linewidth]{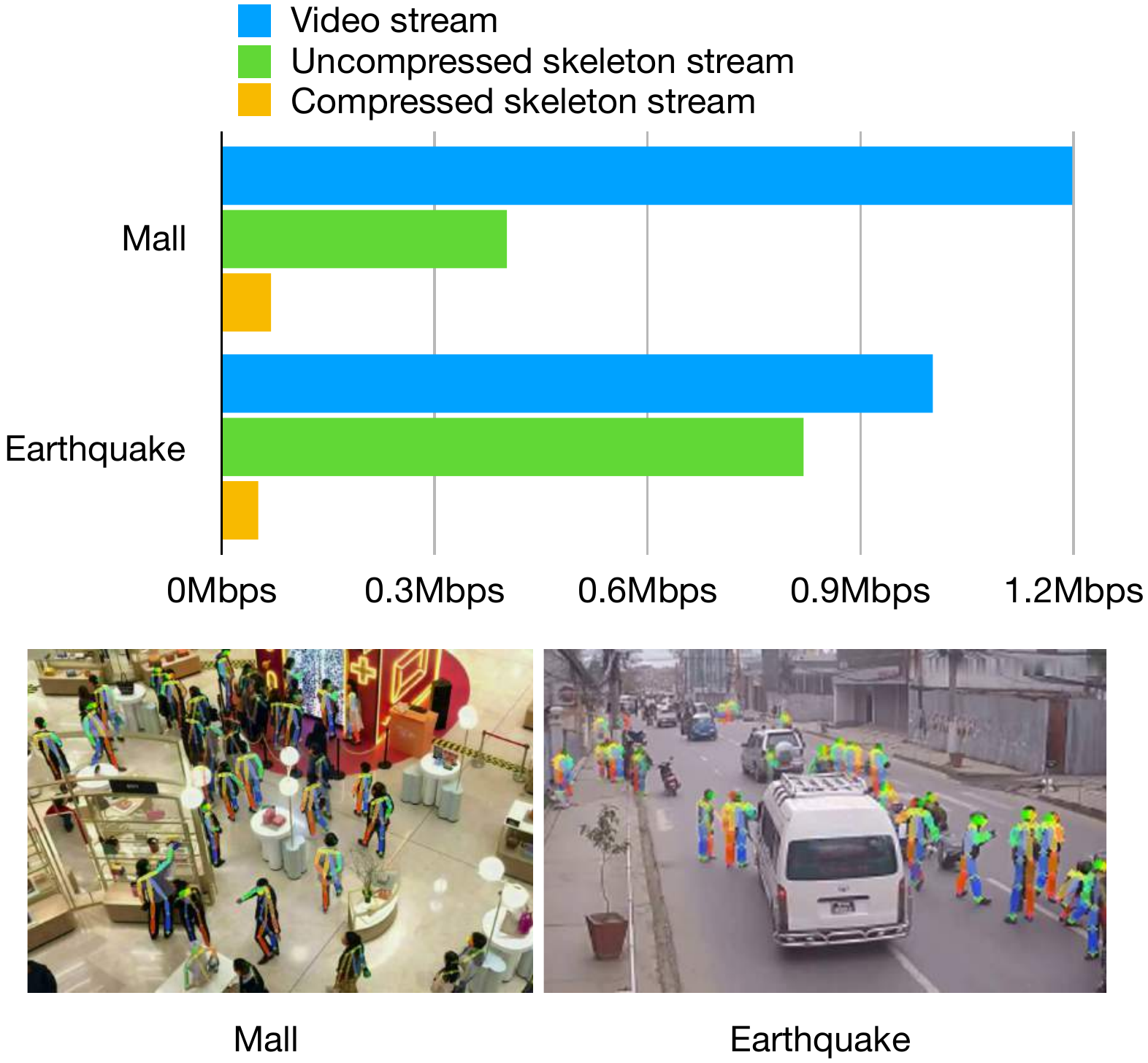}}
\caption{Two typical surveillance video sequences along with uncompressed and compressed skeleton streams. \hxy{Best viewed in color.}}\label{fig:cmp}
\end{figure}

Key-point sequences consist of the coordinates of key points in each frame and the corresponding tracking IDs. With the advances in multimedia systems, such semantic data become non-negligible for complex surveillance scenes with a large number of objects. Figure~\ref{fig:cmp} shows that uncompressed skeleton streams still takes up a costly portion of typical video streams. Therefore, there is an urgency to compress these sequences effectively.

In this paper, we propose a new framework for lossless compression of key-point sequences in surveillance videos to eliminate their spatial and temporal redundancies. The spatial redundancy is caused by correlations of spatial positions, while the temporal redundancy arose from the significant similarities between the positions of object key-points in consecutive frames. The proposed framework as a proposal for key-point compression, has been accepted by the vision feature coding (VFC) group of AITISA\footnote{\href{http://www.aitisa.org.cn/}{http://www.aitisa.org.cn/}} as coding standard for vision features.

We start with a brief review of the feature representation for video analysis, particularly on how key-point information is extracted from videos to generate key-point sequences.  
Consequently, we propose a lossless compression framework for key-point sequences with adaptive selection of prediction modes to minimize spatial and temporal redundancies.
Finally, we present experimental results to showcase the strengths of the proposed framework on various key-point sequences.

\section{FEATURE REPRESENTATION IN EVENT ANALYSIS}
In this section, we discuss several widely used feature representations for event analysis.

\subsection{Digital Video}
As a prevailing representation of video signals, digital videos consist of multiple frames of pixels with three color components. 
Digital video contents can be shown on mobile devices, desktop computers and television.
Compression of digital videos has been well addressed in various studies. The High Efficiency Video Coding (HEVC) standard improves conventional hybrid frameworks like MPEG-2 and H.264/AVC to yield quasi-equivalent visual quality with significantly reduced bit-rates, e.g., 50\% bitrate saving in comparison to H.264/AVC. Recently, the on-going Versatile Video Coding (VVC) standard is expected to further improve HEVC.
\begin{figure*}[!t]
\centering
\subfloat[2D bounding boxes]{\includegraphics[height=0.23\textwidth,width=0.4\textwidth]{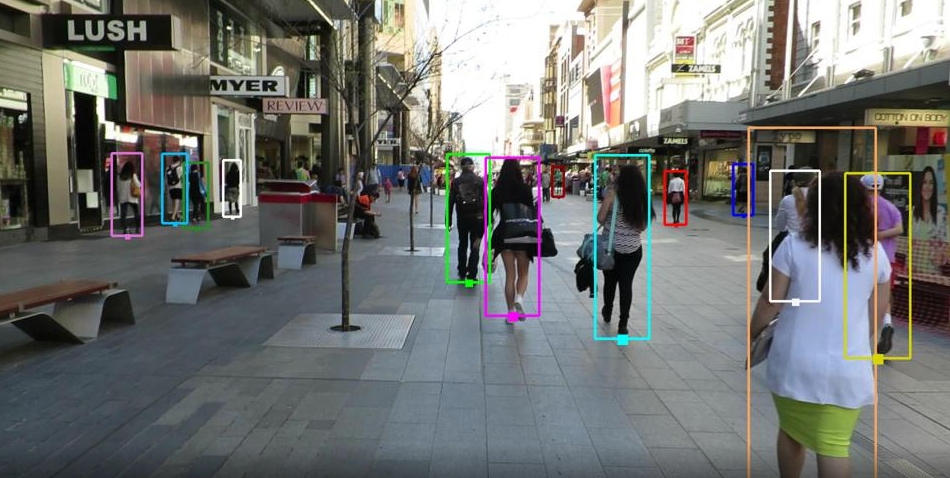}\label{fig:example:2Dbbx}}
\hspace{0.01\textwidth}
\subfloat[3D bounding boxes]{\includegraphics[height=0.23\textwidth,width=0.4\textwidth]{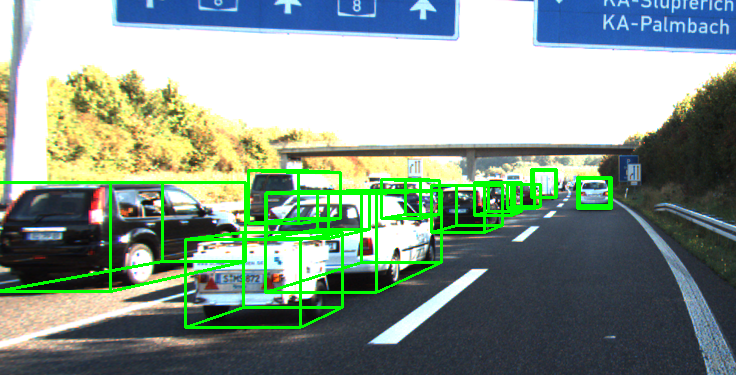}\label{fig:example:3Dbbx}}
\hfill
\vfill
\subfloat[Facial landmarks]{\includegraphics[height=0.23\textwidth,width=0.4\textwidth]{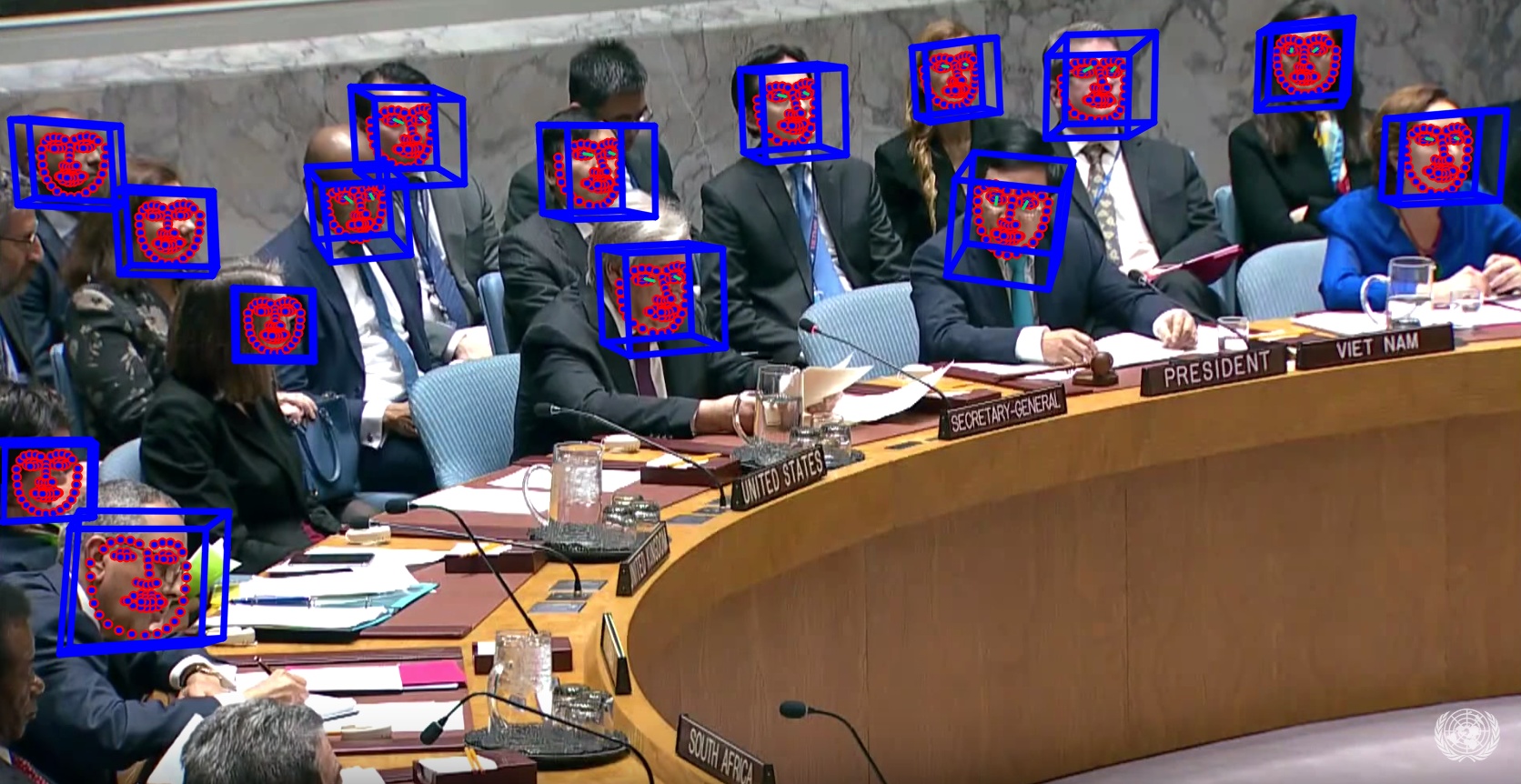}\label{fig:example:face}}
\hspace{0.01\textwidth}
\subfloat[Skeletons]{\includegraphics[height=0.23\textwidth,width=0.4\textwidth]{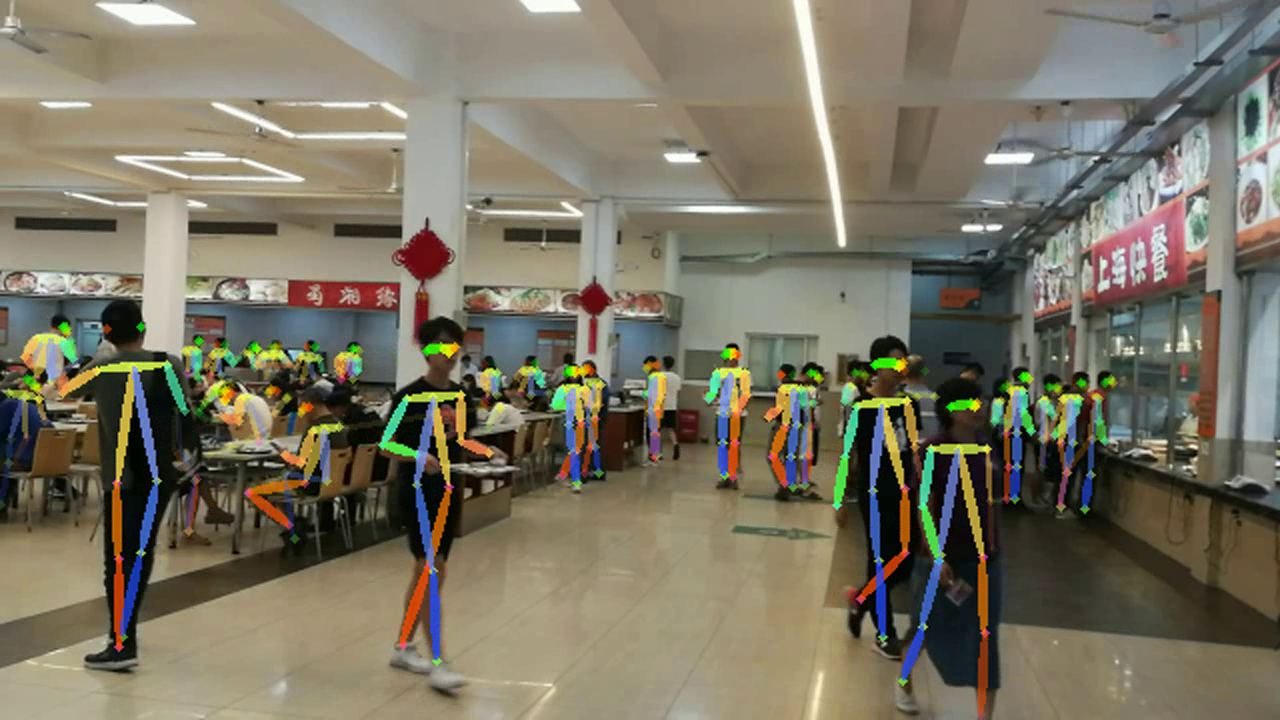}\label{fig:example:ske}}
\hfill
\caption{Examples of different key points. \hxy{Best viewed in color.}}
\label{fig:example}
\end{figure*}

\subsection{Feature Map}
Generally, feature maps (in the form of 4-dimension tensors) are the output of applying filters to the preceding layer in neural networks.
In recent years, deep convolutional neural networks (CNNs) have been utilized to extract deep features for video analysis. These features can be transmitted and deployed to accomplish analysis on the server side. Recently, there has been increasing interest in the compression of deep feature maps. For example, \cite{choi2018deep} employed HEVC to compress quantized 8-bit feature maps.

\subsection{3D Point Cloud}
3D point clouds are popular means of directly representing 3D objects and scenes in applications such as VR/AR, autonomous driving and intelligent transportation systems. They are composed of a set of 3D coordinates and attributes (e.g., colors and normal vectors) for data points in space.
However, communication of point clouds is challenging due to their huge volume of data, which necessitates effective compression techniques.
As such, MPEG is finalizing the standard for point cloud compression that includes both lossless and lossy compression methods.
Typically, the coordinates and attributes of point clouds are compressed separately. Coordinates are decomposed into structures such as octrees~\cite{elseberg2013one} for quantization and encoding. When preprocessed with k-dimensional (k-d) tree and level of details (LoD) description, attributes are compressed with similar encoding process (prediction, transform, quantization, and entropy coding) as traditional image and video coding.

\subsection{Key-Point Sequence}
Various key-point sequences have been considered to improve video representation for urban video analysis. However, costs for transmission and processing are significant as there exist no efficient compression algorithms for key-point sequences.

{\bf 2D Bounding Box Sequence.} A 2D bounding box sequence is a sequence of 2D boxes over time for an object, as shown in Figure~\ref{fig:example:2Dbbx}. A 2D box can be represented by two (diagonal or anti-diagonal) key points. Multiple sequences of 2D boxes can be combined to depict the motion variations and interactions between objects in a scene. As such, these sequences are suitable for human counting, intelligent transportation and autonomous driving.

2D bounding box sequences can be obtained based on object detection~\cite{ren2015faster,law2018cornernet} and tracking~\cite{bewley2016simple}.
2D object detection methods can be classified into anchor free~\cite{law2018cornernet} and anchor based~\cite{ren2015faster} methods. Object tracking~\cite{bewley2016simple} can be viewed as bounding box matching, as it is commonly realized based on a tracking-by-detection strategy.
Furthermore, the \textit{MOT Challenge}~\cite{milan2016mot16} provides a standard benchmark for multiple-object tracking to facilitate the detection and tracking of dense crowds in videos.

{\bf 3D Bounding Box Sequence.} Similar to the 2D case, a 3D bounding box sequence is a sequence of 3D boxes of an object over time. Compared with 2D, 3D bounding boxes offer the size and position of objects in real-world coordinates to perceive their poses and reveal occlusion.
A 3D bounding box shown in Figure~\ref{fig:example:3Dbbx} consists of eight points and can be represented by five parameters. Since an autonomous vehicle requires an accurate perception of its surrounding environment, 3D box sequences are fundamental to autonomous driving systems. A 3D bounding box sequence can be obtained by 3D object detection and tracking methods. 3D object detection can be realized with monocular image, point cloud and fusion based methods~\cite{chen2016monocular,chen2017multi}. Monocular image based methods mainly utilize single RGB images to predict 3D bounding box, but this limits the detection accuracy. Fusion based methods fuse the front-view images and point clouds for robust detection. Tracking with 3D bounding boxes~\cite{frossard2018end} is similar to 2D object tracking, except that modeling object attributes (i.e., motion and appearance) is performed in 3D space. However, uncompressed 3D bounding box sequences are infeasible for transmission.

\begin{figure*}[!t]
\centerline{\includegraphics[width=0.8\textwidth]{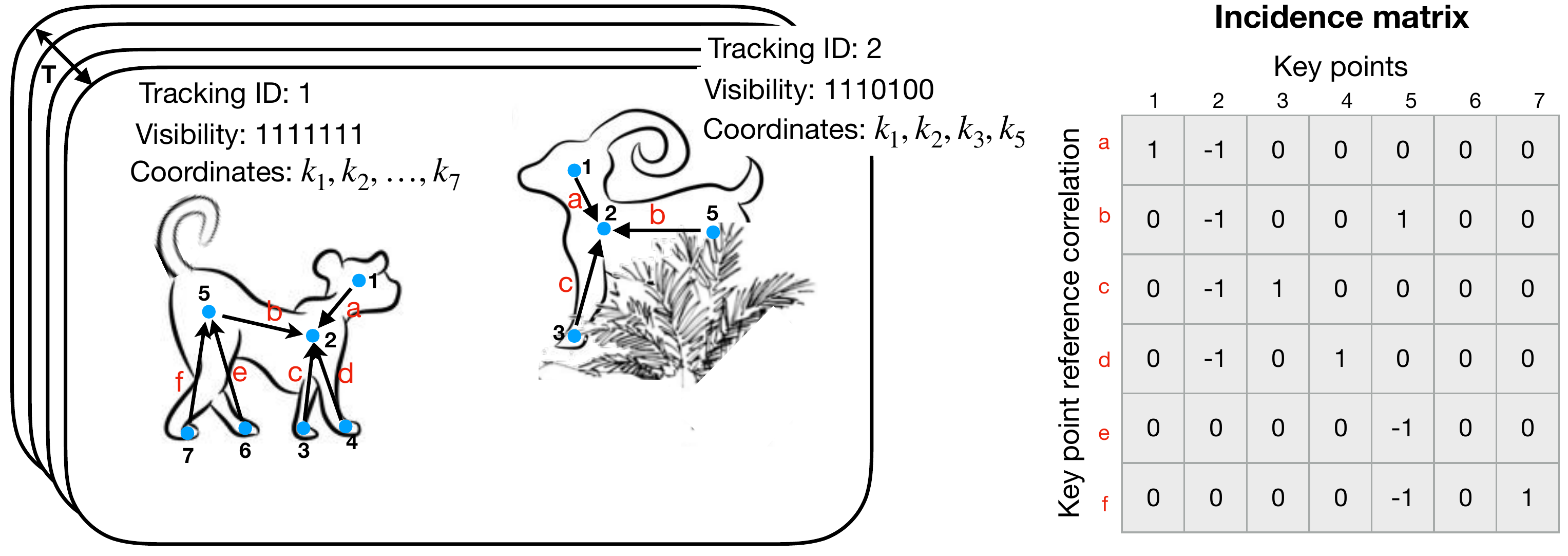}}
\caption{Example of an arbitrary form of key-point sequence in a frame and corresponding incidence matrix. Vertices (key points) are annotated with numbers 1-7 and edges are annotated with letters a-f. \hxy{Best viewed in color.}}
\label{fig:rep_example}
\end{figure*}

{\bf Skeleton Sequence of Human Bodies.}
Skeleton sequences can address various problems, including action recognition, person re-identification, human counting, abnormal event detection, and surveillance analysis. In general, a skeleton sequence of a human body consists of 15 body joints (shown in Figure~\ref{fig:example:ske}), which
provides camera view-invariant and rich information about human kinematics.
Skeleton sequences of human bodies can be obtained by pose estimation and tracking. OpenPose~\cite{cao2017realtime} is the first real-time multi-person 2D pose estimation approach that achieved high accuracy and real-time performance. AlphaPose~\cite{fang2017rmpe} further presents an improved online pose tracker.
\textit{PoseTrack}~\cite{Andriluka_2018_CVPR} is proposed as a large-scale benchmark for video-based human pose estimation and tracking, where data-driven approaches have been developed to benefit skeleton-based video analysis. 

{\bf Facial Landmark Sequence.} Facial landmark sequence, which consists of facial key-points of a human face in video, is widely used in video-based facial behavior analysis. 
Figure~\ref{fig:example:face} provides an example of facial landmarks, where 68 key-points are annotated for each human face.
The dynamic motions in facial landmark sequences can produce accurate temporal representations of faces.
Studies in facial landmark detection range from traditional generative models~\cite{cootes2001active}
to deep neural network based methods~\cite{sun2013deep}.
In addition, facial landmark tracking has been well-studied under constrained and unconstrained conditions~\cite{yang2015facial,yao2016efficient}. 


\section{REPRESENTATION OF KEY-POINT SEQUENCES}
\subsection{Descriptor}
To encode key-point sequences, we propose to represent key-point information in videos with four components: key point coordinate, incidence matrix of key points, tracking ID and visibility indicator.

{\bf Key Point Coordinate.} Key points of each object is expressed as a set of $N$ coordinates of $D$ dimensions (e.g., 2D and 3D):
\begin{equation}
K = \{k_{1}, k_{2}, \dots, k_{N}\},
\end{equation}
where $k_{i} = (p_{i1}, p_{i2}, \dots, p_{iD})$ with $p_{ij}$ the coordinate in the $j$-th dimension for the $i$-th point.

{\bf Incidence Matrix.}
The encoded points can be used as references to predict the coordinates of current point. To efficiently reduce redundancies, an incidence matrix is introduced to define the references to key points. Thus, the key points of an object can be viewed as vertices of a directed graph. An edge directed from point $1$ to point $2$ indicates that $1$ can be a reference point of $2$. Given a key point, one of its adjacent vertices indicated by the incidence matrix are selected for prediction and compression. This suggests that efficient prediction and compression can be achieved by selecting adjacent vertices with higher correlations as references.

{\bf Tracking ID.} Each object is assigned with a tracking ID when it first appears in the video sequence. Note that tracking ID for the same object does not change within the sequence and new objects are assigned a new tracking ID in increasing arithmetic order.

{\bf Visibility Indicator.} Occlusion tends to appear in dense scenarios. This is commonly due to overlapping movements of different objects, and movements in and out of camera view. Similar to PoseTrack~\cite{Andriluka_2018_CVPR} annotations, we introduce a one-bit flag for each key point to indicate whether it is occluded.
\begin{equation}
V = \{v_{1}, v_{2}, \dots, v_{N}\}, \qquad v_{i} \in \{0, 1\}
\end{equation}

\begin{figure*}[!t]
\centerline{\includegraphics[width=\linewidth]{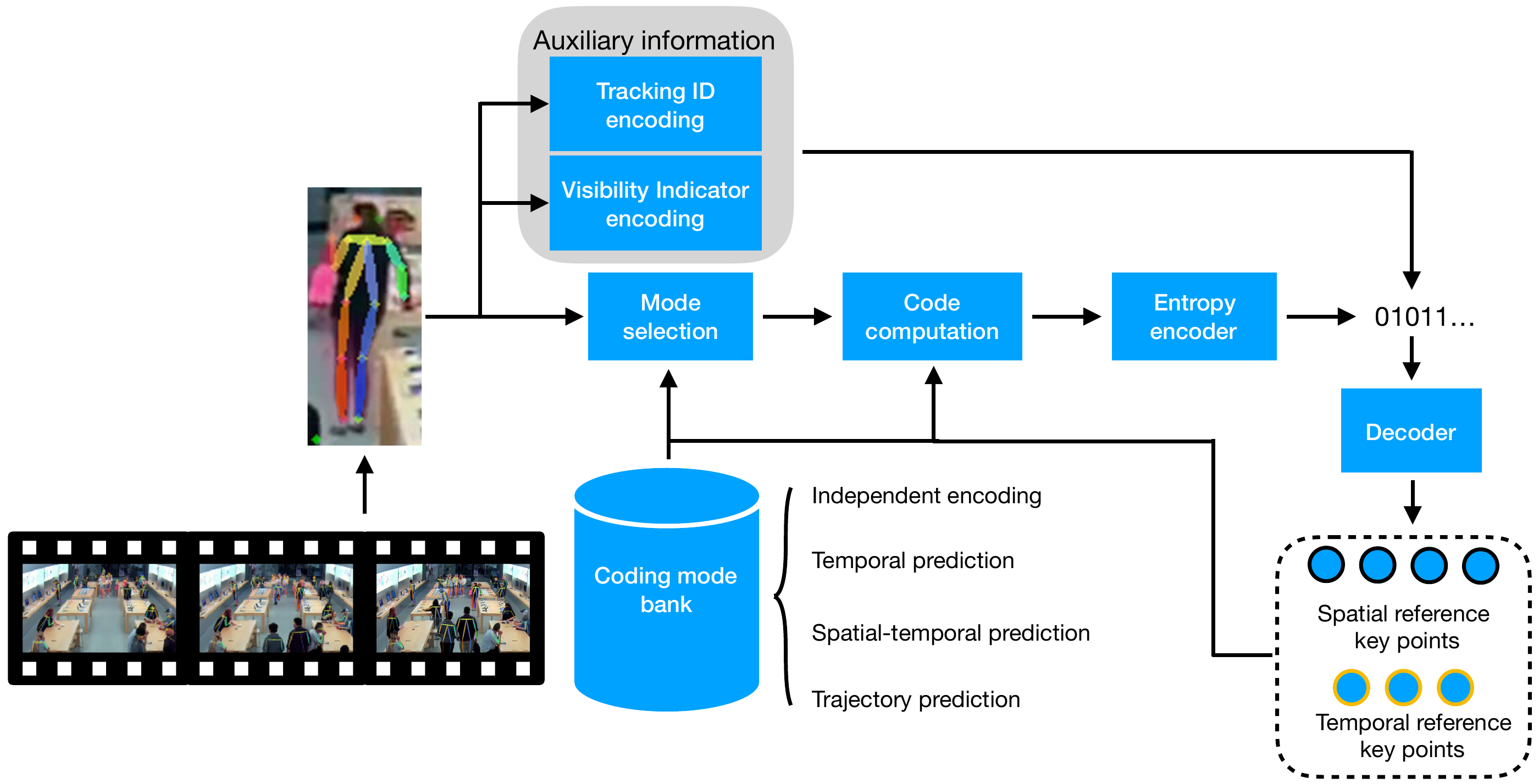}}
\caption{The proposed framework for lossless key-point sequence compression. \hxy{Best viewed in color.}}
\label{fig:framework}
\end{figure*}

\section{LOSSLESS COMPRESSION FOR KEY-POINT SEQUENCES}
\subsection{Framework}
Figure~\ref{fig:framework} illustrates the proposed framework for lossless key-point sequence compression based on the key-point sequence descriptor. Here, we consider to encode the key point coordinates, tracking IDs and visibility indicators, as pre-defined incidence matrices are provided in both encoder and decoder for specific key-point sequences, e.g., facial key points, bounding boxes and skeleton key joints. Similar to H.264/AVC and HEVC, we adopt exponential-Golomb coding to encode prediction residuals.

In this section, four different prediction modes with adaptive mode selection are developed for key-point coordinates, as they consume the bulk of the encoded bitstream.
\hxy{Code computation varies for different encoding modes}. For independent encoding mode, each frame is separately encoded and decoded without reference frames. \hxy{Given references, a predict-and-encode strategy is developed to realize the encoding based on the temporal, spatial-temporal and trajectory prediction modes. Residuals between the original data and their predictions are calculated as the codes to be encoded}. Prediction residuals are then fed into the entropy encoder to generate the bit-stream. It is worth mentioning that prediction modes for the key points can be adaptively predicted using its spatial and temporal neighbors. Furthermore, the predict-and-encode strategy leverages an adaptive prediction method to combine different prediction modes for key-point sequences with various structures and semantic information.
Tracking ID and visibility indicator are also encoded with the auxiliary information encoding module for communication.



\begin{figure*}[!t]
\centering
\subfloat[]{\includegraphics[width=\textwidth]{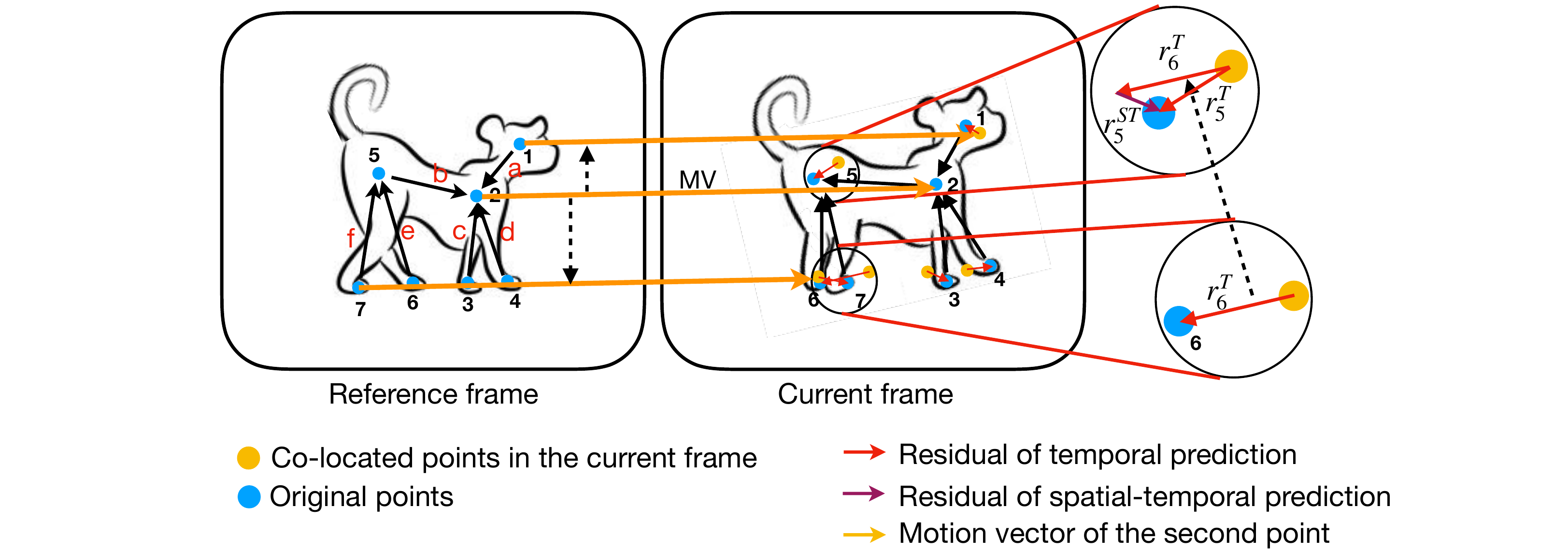}\label{fig:prediction}}
\hfill
\subfloat[]{\includegraphics[width=\textwidth]{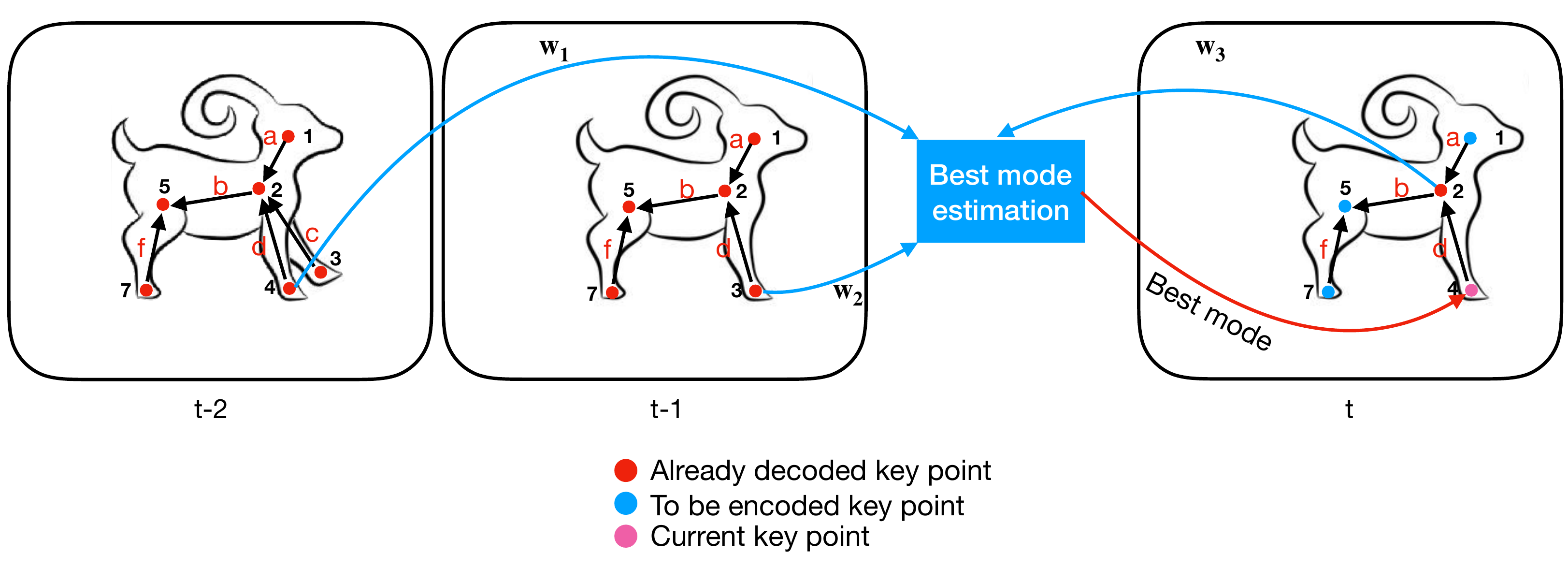}\label{fig:mode_prediction}}
\caption{(a)Illustration of spatial and spatial-temporal prediction modes; (b)Illustration of the best mode estimation with already decoded spatial and temporal references. \hxy{Best viewed in color.}}
\end{figure*}

\subsection{Independent Encoding}
For independent encoding, the key points of a single object are encoded by considering the spatial correlations without introducing references. We first encode the absolute coordinates of the key point $k_{s}$ with zero in-degree. Subsequently, the difference of coordinates between two adjacent vertices defined by the incidence matrix (i.e., the edges) is encoded.
The residual of independent encoding \hxy{$r^{IE}_{i,j}$} between the $i$-th and $j$-th vertices is computed by
\begin{equation}
\hxy{r^{IE}_{i,j}} = k_{i} - k_{j}.
\end{equation}

\subsection{Reference-Based Prediction Modes}
Besides independent encoding, three additional prediction modes are developed for temporal prediction to minimize the residuals with temporal references.

{\bf Temporal Prediction.}
For each object, the correlations between consecutive frames are characterized by the movements, including the translation of the main body and twists of some parts. As shown in Figure~\ref{fig:prediction}, \hxy{we first obtain a co-located prediction (yellow points in the current frame) of the point from the reference frame by motion compensation with the motion vector of the central key point (yellow vector).} 
Consequently, the temporal prediction can be expressed as
\begin{equation}
p_{i}^{t} = k_{i}^{t-1} + MV_{c},
\end{equation}
where $MV_{c} = k_{c}^{t} - k_{c}^{t-1}$ and $k_{c}$ is the key point with maximum out-degree in the incidence matrix. The residuals of temporal prediction \hxy{$r_{i}^{T,t}$} (red dashed vectors) are computed for transmission and reconstruction in a lossless fashion:
\begin{equation}
\hxy{r_{i}^{T,t}} = k_{i}^{t} -  p_{i}^{t}
\end{equation}
However, temporal prediction would be affected by possible twists, i.e, the gap between co-located (yellow and blue) points in the current frame.

{\bf Spatial-temporal Prediction.} The spatial-temporal correlations between key points can be utilized to improve the accuracy of prediction and further reduce the redundancy. Since adjacent points in the incidence matrix are highly correlated in the spatial domain, their movements are probably in the same direction and even with the same distance.
Thus, the redundancy can be further reduced by encoding the residual of prediction $p_{i}$ with respect to the prediction $p_{r(i)}$ of its reference point, as their temporal predictions are very close. For example, as shown in Figure~\ref{fig:prediction}, the spatial-temporal prediction of the 5th point is obtained with the encoded residual of the 6th point (red vector) and the co-located temporal prediction (5th yellow point). \hxy{In this case, we can see that the to be transmitted spatial-temporal residual of the 5th point (maroon vector, $r_{5}^{ST}$) is smaller than the residual $r_{5}^{T}$. Formally, $r_{i}^{ST,t}$ can be computed by}:
\begin{equation}
\label{eq:spatial-temporal}
\hxy{r_{i}^{ST,t}} = p_{i}^{t} - p_{r(i)}^{t},
\end{equation}
where $r(i)$ is the index of $i$-th point's reference. Equation~\ref{eq:spatial-temporal} is equivalent to predicting using $MV_{c}$ and the encoded residual \hxy{$r_{c(i)}^{T,t}$} of the reference point.

{\bf Trajectory Prediction.} The above two modes utilize the MV of the central point to accomplish temporal prediction. However, the motions of different parts of an object are complex, as they vary in direction and distance. Thus, the required bits for coding can be further reduced with more accurate prediction.
For example, when we assume the motion of an object is uniform in a short time (e.g., three frames), the motion from the $(t-1)$-th frame to the $t$-th frame can be approximated with that from the $(t-2)$-th frame to $(t-1)$-th frame. Its predicted value is
\begin{equation}
tp_{i}^{t} = k_{i}^{t-1} + (k_{i}^{t-1} - k_{i}^{t-2}).
\end{equation}
The residual between the predicted value and actual value is computed and transmitted.

The accuracy of trajectory prediction methods can be improved by incorporating more features at the cost of further complexity~\cite{minguez2018pedestrian}. In this paper, we propose a simple and efficient linear prediction based on the previous two frames. 

\subsection{Adaptive Mode Selection}
Independent encoding mode is adopted, when key points are in the first frame or appear for the first time in sequences. When temporal references are introduced, adaptive mode selection is developed for the candidate temporal, spatial-temporal and trajectory prediction modes. The prediction mode \hxy{$m^{\star}$} is estimated from the encoded spatial and temporal reference points with weighted voting:
\begin{equation}
\label{eq:pm}
\hxy{m^{\star}} = \arg\min_{m} \sum_{n \in N_{t}} w_{n}^{t} \times b_{n}^{m} + \sum_{n \in N_{s}} w_{n}^{s} \times b_{n}^{m},
\end{equation}
where $N_{t}$ and $N_{s}$ are the sets of spatial and temporal reference points, $w_{n}^{t}$ and $w_{n}^{s}$ are the weights of the corresponding point $n$ in $N_{t}$ and $N_{s}$, $m$ is the candidate modes for temporal, spatial-temporal prediction and trajectory prediction and $b_{n}^{m}$ is the bit-length of point $n$ encoded with $m$.
As depicted in Figure~\ref{fig:mode_prediction}, the prediction mode of 4th key point in the $t$-th frame is estimated with the reconstructed 4th key point in $(t-2)$-th and $(t-1)$-th frames, along with the encoded first key point in the $t$-th frame with weights $w_{1}$, $w_{2}$, $w_{3}$, respectively.

Equation~\ref{eq:pm} indicates that \hxy{$m^{\star}$} is determined to minimize the average bit-length of its spatial and temporal reference points encoded with all candidate modes. The weights are hyper-parameters that commonly decrease with the growth of the distance between the current point and its neighbors. Note that trajectory prediction will not always be enabled. For example, the object or point exists in the $t$-th and $(t-1)$-th frame would not appear in the $(t-2)$-th frame. \hxy{It is symmetric for the encoder and decoder to determine whether the trajectory prediction is adopted.} Thus, we exclude it from the candidate modes, when unavailable.



\subsection{Auxiliary Information Encoding}
In addition to the key points, tracking ID and visibility indicator are encoded as auxiliary information. Note that they actually consume minimal bit-rates in the output bitstream.

{\bf Tracking ID.} A tracking ID is assigned in arithmetic order to each object when it first appears in the video. For each frame, we sort the objects in ascending order (of tracking IDs) and encode the differences between neighboring tracking IDs.

\begin{table*}[!t]
\renewcommand{\arraystretch}{1.5}
\caption{Average bits for encoding one point and compression ratio for different encoding methods.}
\label{tab:result}
\begin{tabular}{cp{1.2cm}p{1.5cm}p{1.5cm}p{1.5cm}p{1.5cm}p{1.5cm}}
\hline
& Fixed \newline bit-length \newline coding & Independent \newline encoding & Temporal \newline prediction & Spatial-temporal \newline prediction & Trajectories \newline prediction & Multimodal \newline coding \\
\hline
MOT17 & 37.41 & 36.65 (97.97\%) & 14.77 (39.49\%) & 14.77 (39.49\%) & \textbf{13.34 (35.67\%)} & 14.73 (39.37\%) \\
Crowd-event BBX & 38.10 & 36.58 (96.01\%) & \textbf{10.54 (27.66\%)} & \textbf{10.54 (27.66\%)} & 11.31 (29.67\%) & 10.90 (28.59\%) \\
\hline
PoseTrack & 33.35 & 23.30 (69.86\%) & 13.84 (41.50\%) & 13.35 (40.02\%) & 13.37 (40.08\%) & \textbf{12.80 (38.38\%)} \\
Crowd-event skeleton & 33.79 & 14.40 (42.62\%) & 3.17 \newline (9.37\%) & 3.01 \newline (8.92\%) & 4.06 (12.02\%) & \textbf{2.46 \newline (7.27\%)} \\
\hline
nuScenes & 50.29 & 35.48 (70.55\%) & 28.25 (56.18\%) & 27.92 (55.53\%) & 30.78 (61.22\%) & \textbf{27.85 (55.38\%)} \\
\hline
Facial landmarks & 33.11 & 10.20 (30.80\%) & 9.37 (28.31\%) &9.33 (28.18\%) & 9.49 (28.67\%) &\textbf{9.23 (27.87\%)} \\
\hline
\end{tabular}
\end{table*}

{\bf Visibility indicator.} Since visibility indicator changes slowly within two consecutive frames, one bit is used to represent whether it changes for an object. If not, the difference is encoded and transmitted.

\section{EXPERIMENTS}
\subsection{Evaluation Framework}
To demonstrate the robustness of the proposed lossless compression method for key-point sequences, we evaluate four types of key points: 2D bounding boxes, human skeletons, 3D bounding boxes and facial landmarks.

{\bf 2D Bounding Box Dataset.} MOT17 dataset~\cite{milan2016mot16} consists of 14 different sequences (7 training, 7 test sequences). Here, we evaluate the training sequences with ground-truths. We also adopt another important dataset for 2D bounding boxes, i.e., crowd-event BBX dataset~\cite{crowdevent}, which we have constructed. This dataset includes annotated 2D bounding boxes (and corresponding tracking information) in crowed scenes.

{\bf Human Skeleton Dataset.} Two datasets are used for human skeleton compression: (1) PoseTrack; (2) Our crowd-event skeleton dataset~\cite{crowdevent}. For human pose estimation and tracking, PoseTrack~\cite{Andriluka_2018_CVPR} is one of the most widely used dataset with over 1,356 video sequences. Five challenging sequences that contain 7-12 skeletons are chosen as test sequences in this paper.
In our own collected crowd-event skeleton dataset, 
each skeleton is labeled with 15 key joints (e.g., eyes, nose, neck), as shown in Figure~\ref{fig:example:ske}. Compared with the PoseTrack sequences, our crowd-event skeleton dataset contains a larger number of smaller skeletons in crowded scenes.

{\bf nuScenes Dataset.} The nuScenes dataset~\cite{nuscenes2019} is a large-scale public dataset for autonomous driving. It contains 1,000 driving scenes (a 20-second clip is selected for each scene) while accurate 3D bounding boxes sampled at 2Hz over the entire dataset are annotated.

{\bf Facial Landmark Dataset.} We collect three video sequences and label the landmark sequences, as existing facial landmark datasets rarely contain tracking information. The sequences contain 11 to 34 visible human faces, each having about 100 frames on average.

The compression performance is evaluated in terms of (1) average bits for encoding one point (i.e., the ratio between total required bits for encoding and the number of encoded key points) (2) compression ratio (i.e., the ratio between data amount before and after compression). In this paper, the size of uncompressed data is calculated by encoding each coordinate of each key point with a 16-bit universal code, e.g., 32 bits for 2D coordinates and 48 bits for 3D coordinates of each key point. In Tables~\ref{tab:result} and \ref{tab:result2}, the average bits for fixed bit-length coding are obtained by summing up the bit-lengths assigned for coordinates and required for encoding auxiliary information like tracking IDs and visibility indicators.

\begin{table*}
\renewcommand{\arraystretch}{1.5}
\caption{Average bits for encoding one point and compression ratio for different encoding methods with different frame skip scenarios, Gaussian noise level (standard deviation) and data sources.}
\label{tab:result2}
\begin{tabular}{ccp{0.7cm}p{0.7cm}p{1.1cm}p{1.3cm}p{1.1cm}p{1.1cm}p{1.2cm}p{1.1cm}}
\hline
 & Groundtruth?& Frame \newline skip & Noise \newline level &Fixed \newline bit-length \newline coding & Independent \newline encoding & Temporal  \newline prediction & Spatial-temporal  \newline prediction & Trajectories  \newline prediction & Multimodal  \newline coding \\
\hline
\multirow{3}{*}{MOT17} & \ding{51} & 0 & 0 & 37.41 & 36.65 (97.97\%) & 14.77 (39.49\%) & 14.77 (39.49\%) & \textbf{13.34 (35.67\%)} & 14.73 (39.37\%) \\
& \ding{51} & 1 & 0 & 37.41 & 36.66 (97.99\%) & 18.11 (48.40\%) & 18.11 (48.40\%) & \textbf{17.31 (46.26\%)} & 18.29 (48.90\%) \\
& \ding{51} & 2 & 0 & 37.41 & 36.67 (98.01\%) & \textbf{20.29 (54.24\%)} & \textbf{20.29 (54.24\%)} & 20.49 (54.77\%) & 20.67 (55.25\%) \\
& \ding{51} & 0 & 2 & 37.41 & 36.66 (98.00\%) & \textbf{18.56 (49.60\%)} & \textbf{18.56 (49.60\%)} & 19.07 (50.97\%) & 18.87 (50.45\%) \\
& \ding{51} & 0 & 5 & 37.41 & 36.63 (97.91\%) & \textbf{21.78 (58.23\%)} & \textbf{21.78 (58.23\%)} & 22.76 (60.84\%) & 22.07 (58.98\%) \\
& \ding{55} & 0 & 0 & 37.41 & 35.16 (93.99\%) & 15.46 (41.32\%) & 15.46 (41.32\%) & \textbf{14.94 (39.94\%)} & 15.75 (42.11\%) \\

\hline
\hline
\multirow{3}{*}{PoseTrack} & \ding{51} & 0 & 0 & 33.35 & 23.30 (69.86\%) & 13.84 (41.50\%) & 13.35 (40.02\%) & 13.37 (40.08\%) & \textbf{12.80 (38.38\%)} \\
& \ding{51} & 1 & 0 & 33.35 & 23.26 (69.75\%) & 16.76 (50.25\%) & 15.97 (47.89\%) & 17.60 (52.78\%) & \textbf{15.92 (47.74\%)} \\
& \ding{51} & 2 & 0 & 33.35 & 23.23 (69.65\%) & 18.64 (55.89\%) & \textbf{17.67 (52.99\%)} & 20.16 (60.46\%) & \textbf{17.67 (52.99\%)} \\
& \ding{51} & 0 & 2 & 33.35 & 23.33 (69.96\%) & 15.04 (45.10\%) & 14.62 (43.85\%) & 15.27 (45.78\%) & \textbf{14.38 (43.11\%)} \\
& \ding{51} & 0 & 5 & 33.35 & 23.43 (70.25\%) & 17.05 (51.13\%) & 16.79 (50.34\%) & 17.52 (52.52\%) & \textbf{16.50 (49.48\%)} \\
& \ding{55} & 0 & 0 & 33.35 & 22.35 (67.01\%) & 19.14 (57.38\%) & 19.05 (57.12\%) & 19.38 (58.10\%) & \textbf{18.63 (55.86\%)} \\

\hline
\end{tabular}
\end{table*}

\subsection{Results}
Table~\ref{tab:result} reports the performance of different prediction modes. Independent encoding mode is suitable for objects with dense key points (e.g., facial landmark sequences) by exploiting spatial correlations. However, it is inferior to prediction modes based on temporal reference.

The spatial-temporal prediction mode is competitive or slightly better than the temporal prediction mode, due to obvious correlations between spatially adjacent points. The largest performance gap is achieved on PoseTrack, as sports scenes in PoseTrack are regular and predictable. The trajectory prediction mode outperforms other modes on sequences with simple, predictable motions. Consequently, the multimodal coding method is developed to combine different prediction modes and improve compression performance for complex scenes.

The multimodal coding method yields the best average performance on most sequences, which validates the advantages of the proposed scheme.
For 2D bounding box sequences, the multimodal coding method is equivalent or slightly inferior to the single prediction mode based on temporal references. This fact implies that the multimodal coding method is more suited for sequences with complex and unpredictable motions, while the reference-based prediction mode would favor key-point sequences with simple and predictable motions, e.g. 2D bounding box sequences.

We further down-sample the video sequences for evaluations under various motion search ranges. A number of frames are skipped after each frame during encoding. To validate the effectiveness of our approach in real-world applications, we also conduct experiments on data estimated by existing algorithms and noisy data by adding zero-mean Gaussian noise, where a lot of missing and off-target key points exist. Two benchmark datasets (MOT17 and PoseTrack) are evaluated.
Table~\ref{tab:result2} shows that compression performance drops when the frame skipping range increases. More importantly, under different settings, the multimodal coding method still achieves the best performance on all skeleton sequences. It demonstrates the robustness of our proposed scheme.

\section{CONCLUSION AND OUTLOOK}

In this paper, we highlight the problem of lossless compression of features and shown its importance in modern urban computing applications. Importantly, we introduce a lossless key-point sequence compression approach where both reference-free and reference-based modes are presented. Furthermore, an adaptive mode selection scheme is proposed to deal with a variety of scenarios, i.e., camera scenes, key-point sequences and motion degree.Forward looking, we expect that key-point sequence compression methods will play an important role in the transmission and storage of key-point data in urban computing and intelligent analysis.
\bibliographystyle{IEEEtran}
\bibliography{IEEEabrv,refs.bib}
\balance
\begin{IEEEbiography}{Weiyao Lin}{\,}is currently a Full Professor with the Department of Electronic Enigeering, Shanghai Jiao Tong University, Shanghai, China. He received the Ph. D degree from the University of Washington, Seattle, USA in 2010. He served as an associate editor for a number of journals including TIP, TCSVT, and TITS. His research interest includes urban computing and multimedia processing. Contact him at wylin@sjtu.edu.cn.
\end{IEEEbiography}

\begin{IEEEbiography}{Xiaoyi He}{\,}focuses his current research interests on large-scale video compression and semantic information coding. He received the B.S. degree in Electronic Engineering from Shanghai Jiao Tong University (SJTU), Shanghai, China, in 2017. He is currently working toward the M. S. degree at SJTU. Contact him at 515974418@sjtu.edu.cn.
\end{IEEEbiography}

\begin{IEEEbiography}{Wenrui Dai}{\,}is currently an Associate Professor with the Department of Computer Science and Engineering, Shanghai Jiao Tong University (SJTU), Shanghai, China. He received the Ph. D degree from SJTU in 2014. His research interests include learning-based image/video coding, image/signal processing and predictive modeling. Contact him at daiwenrui@sjtu.edu.cn.
\end{IEEEbiography}

\begin{IEEEbiography}{John See}{\,}is a Senior Lecturer with the Faculty of Computing and Informatics at Multimedia University, Malaysia. He is currently the Chair of the Centre for Visual Computing (CVC) and he leads the Visual Processing (ViPr) Lab. From 2018, he is also a Visiting Research Fellow at Shanghai Jiao Tong University (SJTU). Contact him at johnsee@mmu.edu.my.
\end{IEEEbiography}

\begin{IEEEbiography}{Tushar Shinde}{\,}focuses his current research interests on multimedia processing and  predictive coding. He received the M.S. degree in Electrical Engineering from Indian Institute of Technology, Jodhpur (IITJ), India. He is currently working toward the Ph. D degree at IITJ. Contact him at shinde.1@iitj.ac.in.
\end{IEEEbiography}

\begin{IEEEbiography}{Hongkai Xiong}{\,}is a Distinguished Professor in both the Department of Electronic Engineering and the Department of Computer Science and Engineering, Shanghai Jiao Tong University (SJTU). Currently, he is the Vice Dean of Zhiyuan College in SJTU. He received the Ph.D. degree from SJTU in 2003. His research interests include multimedia signal processing and coding. Contact him at xionghongkai@sjtu.edu.cn.
\end{IEEEbiography}

\begin{IEEEbiography}{Lingyu Duan }{\,}is currently a Full Professor with the National Engineering Laboratory of Video Technology, School of Electronics Engineering and Computer Science, Peking University (PKU), Beijing, China. He was the Associate Director of the Rapid-Rich Object Search Laboratory, a joint lab between Nanyang Technological University, Singapore, and PKU, since 2012. Contact him at lingyu@pku.edu.cn.
\end{IEEEbiography}

\end{document}